\documentclass[12pt,a4paper]{article}

\usepackage{rotating}
\usepackage{graphicx}
\usepackage{amsmath,amssymb}
\usepackage{chapterbib,cite,afterpage,float}

\usepackage{color}

\definecolor{response}{rgb}{0,0,0}

\usepackage{times}

\newcounter{lastnote}

\author{Daniel Silk$^{1,2}$, Paul D. W. Kirk$^{1,2}$, Chris P. Barnes$^{1,2}$, Tina Toni$^{1,2,3,\dagger}$, \\Anna Rose$^{3,4}$, Simon Moon$^3$,  Margaret J. Dallman$^{3,4}$ \\ \& Michael P.H. Stumpf$^{1,2,3,5\ast}$}

\title{Designing Attractive Models via Automated Identification of Chaotic and Oscillatory Dynamical Regimes\footnote{The authors would like to dedicate this article to the memory of Jaroslav Stark.}}

\date{}

\begin{document} 


\maketitle

\begin{center}
${^1}$Centre for Bioinformatics, Imperial College London\\
${^2}$Institute of Mathematical Sciences, Imperial College London\\
${^3}$Centre for Integrative Systems Biology at Imperial College London\\ 
${^4}$Division of Cell and Molecular Biology, Imperial College London\\
${^5}$Institute of Chemical Biology, Imperial College London\\

\noindent
{\small $^\ast$ To Whom Correspondence Should be Addressed: m.stumpf@imperial.ac.uk}\\
\noindent
{\small $^\dagger$ Present Address: Department for Biological Engineering, Massachusetts Institute of Technology}

\end{center}


\begin{abstract}
Chaos and oscillations continue to capture the interest of both the scientific and public domains. Yet despite the importance of these qualitative features, most attempts at constructing mathematical models of such phenomena have taken an indirect, quantitative approach, e.g. by fitting models to a finite number of data-points. Here we develop a qualitative inference framework that allows us to both reverse engineer and design systems exhibiting these and other dynamical behaviours by directly specifying the desired characteristics of the underlying dynamical attractor. This change in perspective from quantitative to qualitative dynamics, provides fundamental and new insights into the properties of dynamical systems.
\end{abstract}

Mathematical modelling requires a combination of experimentation, domain knowledge and, at times, a measure of luck. Beyond the intrinsic challenges of describing complex and complicated phenomena, the difficulty resides at a very fundamental level with the diversity of models that could explain a given set of observations. This is a manifestation of the so-called {\em inverse problem}\cite{Tarantola:2005aa}, which is encountered whenever we aim to reconstruct a model of the process from which data have been generated. Exploring the potential space of solutions computationally can be prohibitively expensive and will generally require sophisticated numerical approaches or search heuristics, as well as expert guidance and manual interventions. Parameter estimation\cite{Moles:2003aa}, model inference\cite{Hartemink:2005aa} and model selection\cite{Vyshemirsky:2008p14865,Toni:2010p4095} all address aspects of this problem. 
\par
\afterpage{
\begin{figure}[H]
\includegraphics[width=14cm]{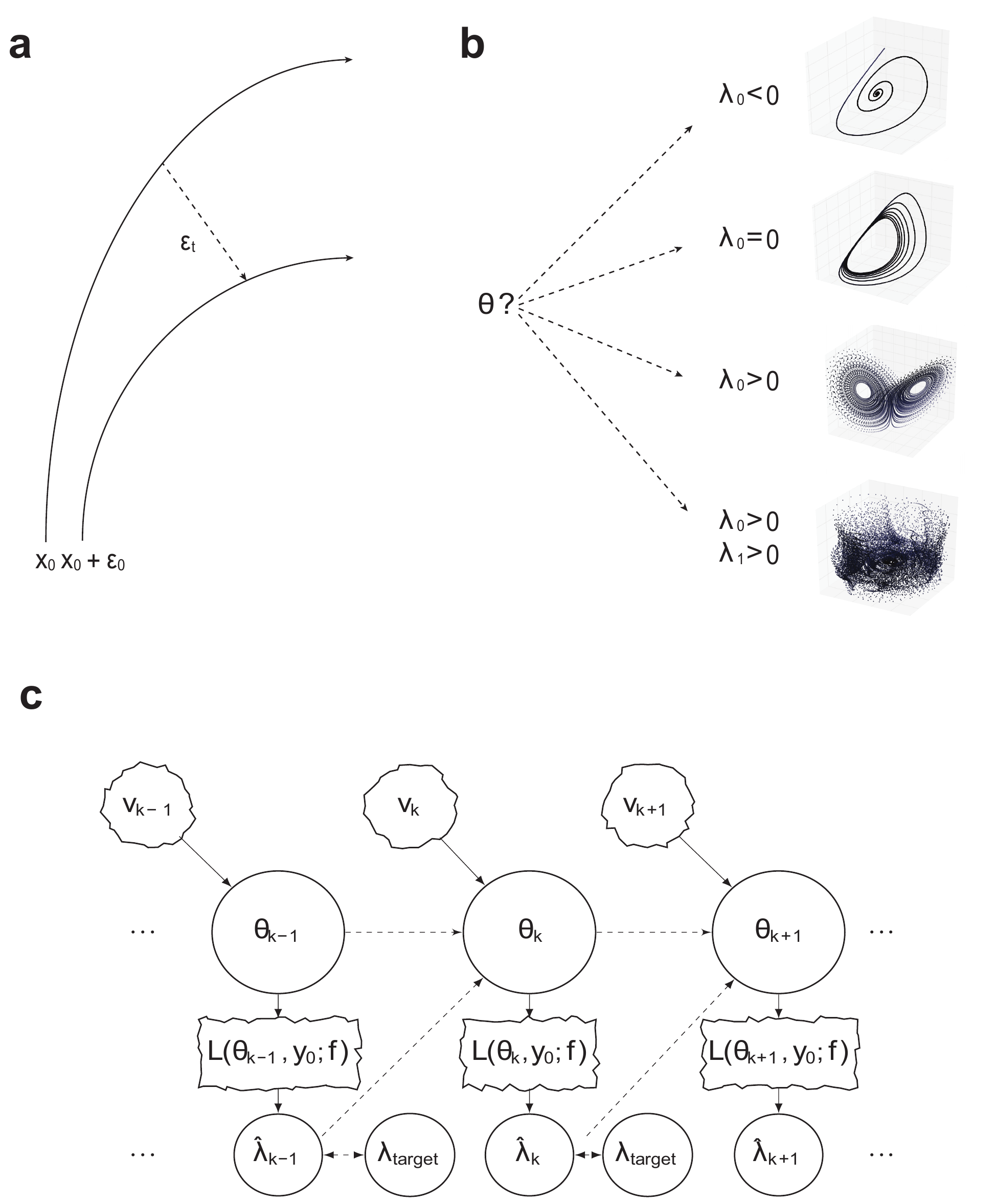}
\caption{\footnotesize{{\color{response}Encoding and inferring the desired dynamics. (a) Lyapunov exponents (LEs), $\lambda_0,...\lambda_n$, characterise the contraction/expansion of an initially small perturbation, $\epsilon_0$, to the system. (b) The leading LE determines the principal dynamics and characteristics of the attractor of a dynamical system. For $\lambda_0<0$ the attractor will be a stable fixed-point; stable oscillating solutions will be obtained if $\lambda_0=0$; for $\lambda_0>0$ we observe chaos and the system will exhibit a so-called strange attractor; if more than one LE is positive, then we speak of hyperchaos and the attractor will exhibit behaviour with similar statistical properties to white noise. (c) Key steps in the unscented Kalman filter (UKF) for qualitative inference. At the $k^{th}$ iteration, the current prior parameter distribution is formed by perturbing the previous posterior, $\theta_k$, with the process noise $v_k$. The distribution of LEs for the model $f$ induced by the prior parameter distribution is calculated via the LE estimation routine L and the unscented transform. Comparing the mean LE, $\hat \lambda_k$, to the target LE, $\lambda_{\text target}$, the prior parameters are updated using the UKF update equations. As the filter proceeds, parameters are found that locally minimise the sum of squared error between target and estimated LEs.}}}
\label{LyapunovExponents}
\end{figure}}
The inverse problem also applies in a different context: the design of systems with specified or desired outputs. Here again we have a multitude of different models --- or for sufficiently complicated models a potentially vast range of parameters --- that fulfil a given set of design objectives. Therefore system design can be fraught with the same challenges as statistical inference or reverse engineering tasks: in the former case we want to learn the existing structure and properties of a system that has produced certain types of data, while in the latter we want to design constructible systems that will reliably and robustly exhibit certain types of behaviour.
\par
These challenges are often further exacerbated by unsuitable or insufficient encoding of the behaviour that we observe (in natural systems) or would like to see (in designed systems). For example, if we aim to estimate parameters describing an oscillating system from a series of observations, then it is possible to get good and even globally optimal fits to the data, without finding a qualitatively acceptable solution. Various methods of qualitative inference have been developed to address this issue; the topology of bifurcation diagrams{\color{response} \cite{Lu:2006p2418,Khinast:1997vx}}, local stability properties of dynamically invariant sets{\color{response}\cite{Chickarmane:2005p4836,Conrad:1984ul}}, {\color{response}symbolic sequences of chaotic systems\cite{Wu:2004td}} and temporal logic constraints\cite{Batt:2007p2419, Rizk:2008p5031} have variously been used to drive parameter searches, or for model checking. However these methods are either limited in the complexity of behaviour they can detect, or by conditioning on surrogate data (e.g. forcing solutions through a small number of points), they suffer in the same way as quantitative approaches. The method proposed here extends the scope of the promising but underdeveloped class of qualitative parameter estimation algorithms\cite{Endler:2009p2387}, allowing detection and control of the most complex and elusive dynamical behaviours, such as oscillations, chaos and hyperchaos.
\par
We consider models of the general form 
\begin{eqnarray}
\frac{d y(t)}{dt}=f(y(t),y_0; \theta),
\end{eqnarray}
where $y(t)$ denotes the n-dimensional state of the system at time $t$, $f$ is the gradient field characterised by a parameter vector, $\theta$, and $y_0=y(0)$ are the initial conditions, which may be unknown, too. Coaxing the solutions of such systems into exhibiting a desired dynamical behaviour is reliant upon the ability to, firstly, encode the behaviour sufficiently as constraints upon a set of model properties that may be conveniently evaluated, and secondly, to identify regions in parameter space for which these constraints are satisfied. {\color{response}Here we meet these challenges using a combination of statistical and dynamical systems techniques. In particular, we pose the problem within a state-space framework, where the observation function corresponds to evaluating the type of attractor exhibited by the model with given parameters and initial conditions. We then exploit the flexibility and efficiency of the unscented Kalman filter (UKF) to systematically move in parameter space until the desired or expected dynamical behaviour is exhibited. The approach, outlined in Fig. 1 and developed fully in the Methods and online Supplementary Information, is demonstrated below within different contexts, covering some classical dynamical model systems and electronic circuits that exhibit oscillations, chaos and hyperchaos, and a biological regulatory system that exhibits oscillatory behaviour.}
\par

\section*{Results}
\afterpage{
\begin{figure}[H]
\begin{center}
\includegraphics[width=15cm]{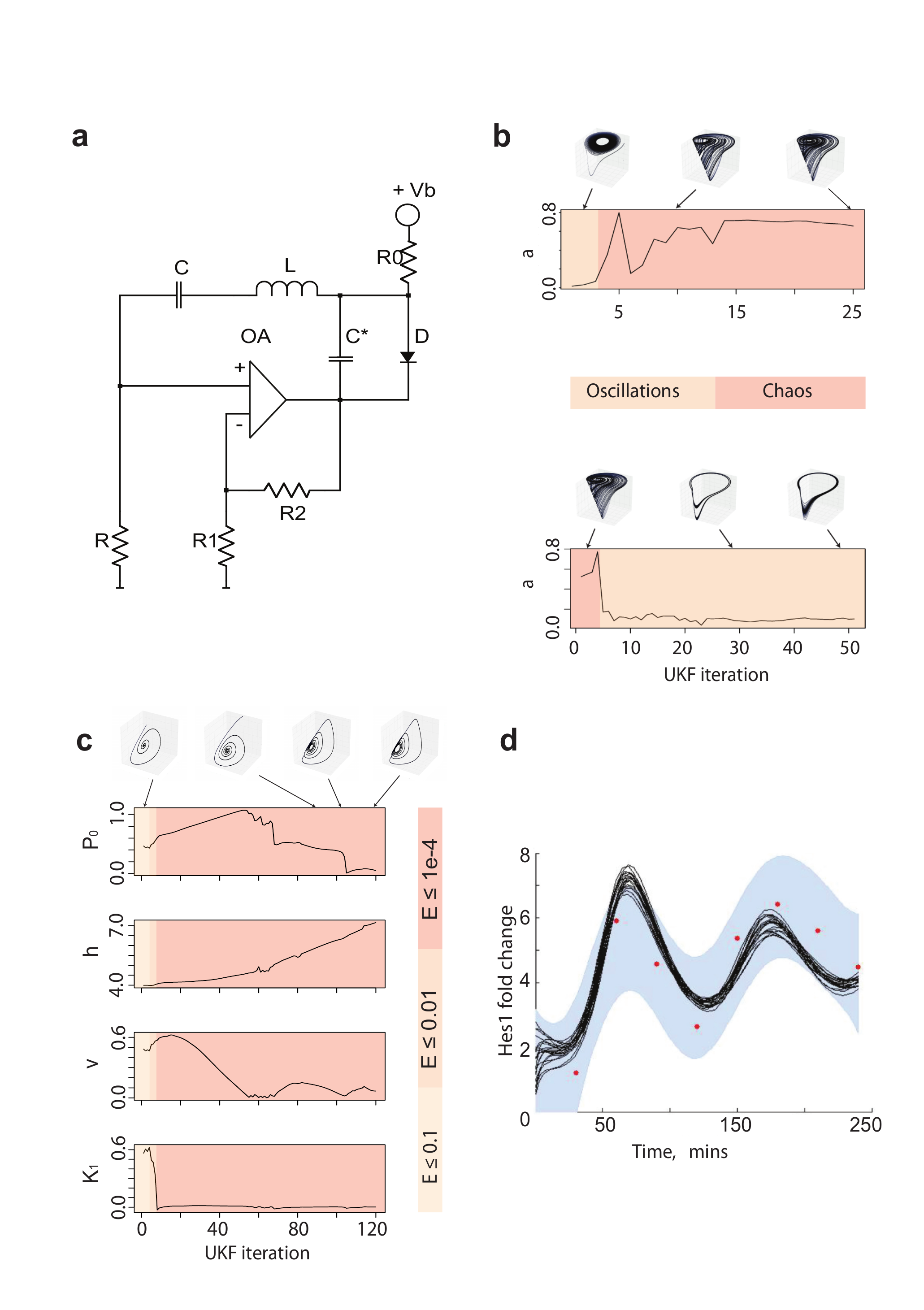}
\end{center}
\caption{\footnotesize{{\color{response}Detecting and controlling chaos and oscillations.  Plots show the estimated parameters at successive iterations of the unscented Kalman filter. Snapshots of the developing attractor are shown above the plots. The sum of squared error (E) is indicated for different sections of the parameter trajectories. (a, b) The filter is able to drive the electric circuit between oscillations and chaos in less than 10 iterations.} (c) Parameter trajectories for a simple model of the Hes1 regulatory system that yield oscillations. Several regions in parameter space can be identified that exhibit oscillatory behaviour. (d) Examples of trajectories generated from a region in parameter space which was found using our approach. Here we used the qualitative inference procedure in order to elicit a prior to be used for parameter inference. Trajectories were sampled from the prior. Data are indicated by red circles and represent fold change in Hes1 mRNA; the blue strips indicate the confidence intervals obtained using Gaussian Process regression, in which standard Gaussian noise is assumed, with maximum marginal likelihood estimates for the other hyperparameters\cite{Kirk:2009aa}. }} 
\label{Hes1}
\end{figure}
}

\subsection*{Oscillations and  chaos in electronic circuits}

The elimination of chaos from a system, or conversely its ``chaotification'', have potential applications to biological, medical, information processing and other technological systems\cite{GuanrongChen:1997p5902}. Here, we use a simple electric circuit\cite{Tamasevicius:2005p3978} (shown in Fig. 2a), to {\color{response} illustrate how our method can be used to tune the system parameters such that the dynamics are driven into and out of chaos.} The circuit model includes a parameter $a$, representing the scaled resistance of a variable resistor, $R_2$, which we make the lone subject of the inference. In turn we start the system in an oscillatory regime and tune the parameter according to the posterior predictions at each step of the UKF, until we enter a chaotic regime, and {\em vice versa} (see Fig. 2b). The two desired behaviours are encoded as constraints only upon the target maximal Lyapunov exponent (LE), specifying, $\lambda_1=0$, for oscillations, and, $\lambda_1= d > \delta_{tol}$ for chaos, where $\delta_{tol}$ is taken larger than the expected error in the  LE estimation procedure, as discussed in the Supplementary Information online.
\par
For systems of this size, the qualitative dynamical regimes can be explored exhaustively and in short time (finding the desired behaviour takes minutes even for moderate sized systems).

\subsection*{Detecting oscillations in immune signalling}
Oscillations appear to be ubiquitous in nature, yet for reasons noted above they often remain elusive to quantitatively driven parameter inference techniques. Here we consider a dynamical system describing the expression levels of the transcription factor Hes1, which is involved in regulating the segmentation of vertebrate embryos\cite{Momiji:2008p5042}. Oscillations of Hes1 expression levels have been observed {\it in vitro} in mouse cell lines, and reproduced using various modelling approaches, including continuous deterministic delay\cite{Monk:2003p5886, Momiji:2008p5042} and discrete stochastic delay models\cite{Barrio:2006p3389}. We investigate a simple three component ODE model of the regulatory dynamics with mRNA transcription modelled by a Hill function,

\begin{eqnarray}
\dot{M} &=&-k_{deg}M+1/(1+(P_2/P_0)^{h})\\
\dot{P_1} &=& -k_{deg}P_1+\nu M - k_1P_1\\
\dot{P_2} &=& -k_{deg}P_2+k_1P_1,
\end{eqnarray}

\noindent where state variables $M$, $P_1$ and $P_2$, are the molecular concentrations of Hes1 mRNA, cytoplasmic and nuclear proteins respectively. The parameter $k_{deg}$ is the Hes1 protein degradation rate which we assume to be the same for both cytoplasmic and nuclear proteins, $k_1$ is the rate of transport of Hes1 protein into the nucleus, $P_0$ is the amount of Hes1 protein in the nucleus when the rate of transcription of Hes1 mRNA is at half its maximal value, $\nu$ is the rate of translation of Hes1 mRNA, and $h$ is the Hill coefficient. For the inference we take, $k_{deg}$, to be the experimentally determined value of $0.03$ $min^{-1}$ \cite{Hirata:2002p4293}. 
\par

In Fig. 2c we show the results for the inference using our algorithm on the model shown above. Note that the value inferred for parameter $k_1$, is significantly lower than the range of values investigated for the continuous deterministic delay model of  H. Momiji and N. A. M. Monk\cite{Monk:2003p5886}. Interestingly, repeating the inference with different initial parameter sets leads to similar values of  $k_1$ ($k_1< 0.01$), but to a broad range of values for the other parameters, all of which result in oscillatory behaviour.  Our qualitative inference thus suggests that oscillations of Hes1 protein and mRNA levels are strongly dependent upon maintaining a low rate of transport of Hes1 protein into the nucleus, and that the dependence on other system parameters is less strong. As $1/k_1$ is the expected time Hes1 spends in the cytoplasm, this corresponds to the delay that had previously been posited to be necessary for such oscillations to occur\cite{Monk:2003p5886}. Our approach readily identifies a parameter regime exhibiting oscillatory dynamics without explicitly requiring (discrete) time-delays.

 \par
 \afterpage{
 \begin{figure}[H]
 \begin{center}
\includegraphics[width=15cm]{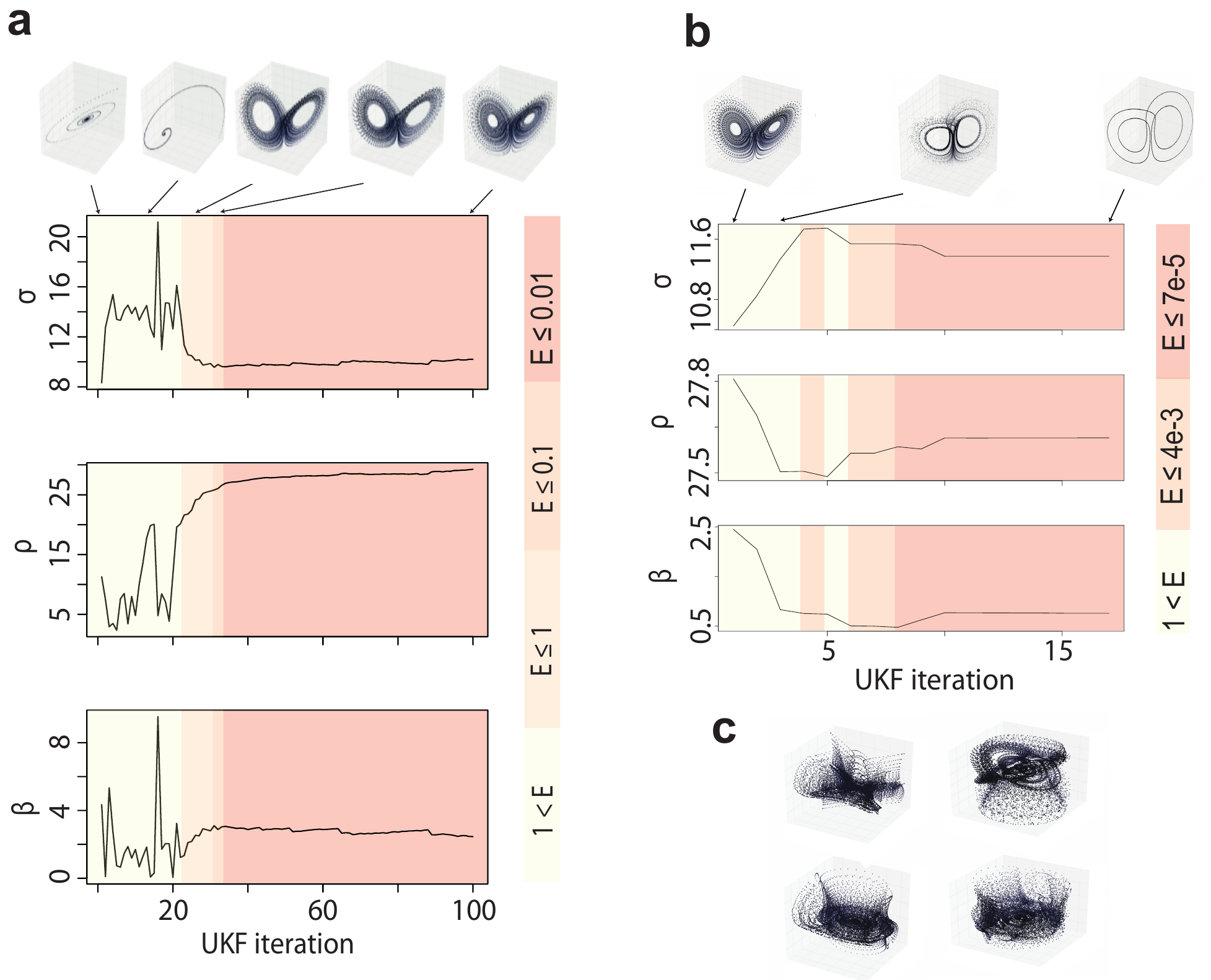}
\end{center}
\caption{\footnotesize{{\color{response}Designing attractive models. (a) Inferring a complete spectrum. After only 22 iterations, the characteristic ``butterfly" strange attractor emerges. The final parameters and LEs are $\sigma=10.2$, $\rho=29.2$, $\beta=2.45$ and $(0.899, 2.74\times 10^{-4}, -14.6)$. (b) A function of the Lyapunov exponents, the Kaplan-Yorke fractal dimension may also be used to specify the desired attractor. Here parameters for a target dimension of 1 are found for the Lorenz system within 20 iterations, giving rise to a limit cycle as required by the theory. }(c) 3-dimensional projections of the hyperchaotic system with parameter vector $(a, b, c, d, e, f)$=$(49.98, 35.86, 30.5, 1.35, 36.6, 33.8)$ and corresponding LEs $(31.8, 16.8, -19.1, -71.4)$. A very chaotic attractor. Within few iterations our algorithm was able to drive the system towards an attractor characterised by Lyapunov exponents twice the size of any that had previously been reported.}} 
\label{Hyperchaos}
\end{figure}}

{\color{response}Next, we used the qualitative inference result as the basis to estimate the model parameters from the Hes1 data described below.  An approximate Bayesian computation algorithm (ABC SMC\cite{Toni:2009p9197}), capable of sampling from non-Gaussian and multimodal posteriors, was employed and Fig. 2d shows the fits of simulated trajectories for 20 parameters drawn randomly from the resulting posterior distribution; these are in good agreement with the confidence intervals (the blue bands in Fig. 2d), which can be obtained from the time-course data via a Bayesian nonparametric method \cite{Kirk:2009aa}. It is worth noting that using the UKF alone, we could in principle consider the Lyapunov exponents and data together in order to infer parameters that are both qualitatively and quantitatively acceptable. However by splitting the inference, we take advantage of the strengths of each algorithm within the Bayesian framework; first we exploit the efficiency of the unscented Kalman filter to work with a sophisticated encoding of the desired behaviour that is computationally expensive to calculate; subsequently we use this qualitative information in order to construct suitable priors for an ABC method capable of dealing with non-Gaussian posteriors.}

\subsection*{Designing attractors}
{\color{response}While the maximal LE alone is sufficient to encode fixed points, limit cycles and strange attractors, we may include additional target exponents to design the complete Lyapunov spectrum (Fig. 3a), design the (Kaplan-Yorke) fractal dimension\cite{Castiglione:2008aa} ($D = k + \sum_{i=1}^{k}\lambda_i/|\lambda_{k+1}|,$ where $k$ is the largest integer for which $ \sum_{i=1}^{k}\lambda_i \geq 0$) of a system's attractor (Fig. 3b), or drive models to behave hyper-chaotically (Fig. 3c). 

The first two of these applications are illustrated with the Lorenz system\cite{Lorenz:1963p4232} which has become a canonical example of how sensitivity to initial conditions can give rise to unpredictable behaviour. The model is known to exhibit a chaotic regime with Lyapunov exponents, $\Lambda= (0.906, 0, -14.57)$, for parameter vector $(\sigma, \rho, \beta)=(10, 28, 8/3)$. Here we infer back these parameters, starting with different prior means, by setting our target Lyapunov spectrum to $\Lambda$. If we restrict the parameter search to the region $[0,30]^3$, as described in the Supplementary Information, we are able to do this reliably from random starting positions. The parameter trajectories and evolving attractor of a representative run of the inference algorithm is shown in Fig. 3a, where the sum of squared error between estimated and target LEs is less than $8\times 10^{-5}$ after the 100th iteration. Without constraints on the parameters the inference algorithm converges to different parameter combinations that display indistinguishable LEs. This allows us to assess (for example) the robustness of chaotic dynamics by mapping systematically the regions of parameter space that yield similar Lyapunov exponents. Fig. 3b shows how the fractal dimension - a function of the Lyapunov exponents - may also be tuned (in this example, halved). While computational difficulties have in the past precluded such investigations, our approach allows us to map attractor structures (and the range of parameters giving rise to similar attractors) very efficiently.}

For the third application of driving a system into hyper-chaos, we investigate a four dimensional system with six parameters, whose significance lies in having two very large LEs ($\lambda_1 \in [10.7741, 12.9798]$ and $\lambda_2 \in [0.4145, 2.6669]$) over a broad parameter range \cite{Qi:2008p586}. The resulting highly complex deterministic dynamics share statistical properties with white noise, making it attractive for engineering applications such as communication encryption and random number generation. By setting large target values of $\lambda_1$ and $\lambda_2$, we use our method to obtain parameters for which the system displays LEs that are over two times bigger than previously found for the system. Fig. 3c, shows the three dimensional projections of the resulting hyper-chaotic attractor. 
\par
These are, of course, toy-applications, but they demonstrate the flexibility and potential uses of this approach: we can really start to explore qualitative behaviour in a numerically efficient and speedy manner. For example, it becomes possible to map (or design) the qualitative characteristics of complex systems and to test robustness of qualitative system features systematically.

\section*{Discussion}
We have demonstrated that it is possible to use statistical inference techniques in order to condition dynamical systems on observed (biological oscillations in Hes1) or desired qualitative characteristics (oscillations, chaos and hyper-chaos in natural and engineered systems).  This provides us with unprecedented ability to probe the workings of dynamical systems. Here we have only used the approach for inference and design of the attractors of dynamical systems, as encoded by their Lyapunov exponents. {\color{response}This, however, has already been enough to show that it is not necessary to impose discrete time-delays in order to explain the oscillations in the Hes1 system \cite{Monk:2003p5886, Momiji:2008p5042,Barrio:2006p3389}.}
\par
A focus on qualitative features has several advantages: first, it is notoriously difficult, for example, to ensure that parameter inference preferentially (let alone exclusively) explores regions in parameter space that correspond to the correct qualitative behaviour such as oscillations. This is the case for optimisation as well as the more sophisticated estimation procedures. Arguably, however, solutions which display the correct qualitative behaviour are more interesting than those which locally minimise some cost-function in light of some limited data. Obviously, in a design setting ensuring the correct qualitative behaviour is equally important. 
\par
Second, the numerical performance of the current approach allows us to study fundamental aspects related to the robustness of qualitative behaviour. This allows us for the first time to ascertain how likely a system it is to produce a given Lyapunov spectrum (and hence attractor dimension) for different parameter values, $\theta$. Our approach, coupled with  means of covering large-dimensional parameter spaces, such as Latin-hypercube or Sobol sampling \cite{Saltelli:2008aa}, allows us to explore such qualitative robustness. Or more specifically, we can map out boundaries separating areas in phase space with different qualitative types of behaviour. We can also drive systems into regions with Lyapunov exponents of magnitudes not previously observed. The last aspect will have particular appeal to information and communication scientists as such hyper-chaos shares important properties with white noise and potential applications in cryptography and coding theory abound \cite{Schroeder:2008aa}.
\par
Finally, our approach can also be used to condition dynamical systems on all manner of observed or desired qualitative dynamics, such as threshold behaviour, bifurcations, robustness, temporal ordering etc.. To rule out that a mathematical model can exhibit a certain dynamical behaviour will, however, require exhaustive numerical sampling of the parameter space; but coupled to ideas from probabilistic computing\cite{Mitzenmacher:2005aa}, our procedure lends itself to such investigations. Both for inference and design problems we foresee vast scope for applying this type of qualitative inference-based modelling.  There is currently still a lack of understanding between the interplay of qualitative and quantitative features of dynamical systems\cite{Kirk:2008p3388}; this becomes more pressing to address as the systems we are considering become more complicated and the data collected more detailed. Flexibility in parameter estimation --- whether based on qualitative or quantitative system features --- will be an important feature for the analysis of such system, as well as the design of synthetic systems in engineering and synthetic biology.

\section*{Methods}

\subsection*{Encoding dynamics through Lyapunov exponents}
Consider a continuous time dynamical system  ---  similar results hold for the discrete case ---  described by,
\begin{eqnarray}
\frac{d{ y_t}}{dt}=f({ y}_t),
\end{eqnarray}
where $f$ is an $n$-dimensional gradient field. {\color{response}To study the sensitivity of $f$ to initial conditions, we consider the evolution of an initially orthonormal axes of $n$ vectors, $\{{ \epsilon}_1, { \epsilon}_2, ..., { \epsilon}_n\}$, in the tangent space at ${ y}_0$. At time $t$, each $\epsilon_i$ satisfies the linear equation,
\begin{eqnarray} \label{perturbationEvolution}
\frac{{d \epsilon}_i}{dt} = Df({ y}_t)\cdot{ \epsilon}_i,
\end{eqnarray}
\noindent where $Df({ y}_t)$ is the Jacobian of $f$ evaluated along the orbit $y_t$. Equations (2) and (3) describe the expansion/contraction of an n-dimensional ellipsoid in the tangent space at $y_t$, and we denote the average exponential rate of growth over all $t$ of the $i$th principal axis of the ellipsoid as $\lambda_i$. The quantities, $\lambda_1 \geqslant \lambda_2 \geqslant ... \geqslant \lambda_n$, are called the global Lyapunov exponents (LEs) of $f$.} In particular, the sign of the maximal LE, $\lambda_1$, determines the fate of almost all small perturbations to the system's state, and consequently, the nature of the underlying dynamical attractor. For $\lambda_1 < 0$, all small perturbations die out and  trajectories that start sufficiently close to each other converge to the same stable fixed point in state-space; for $\lambda_1 = 0$, initially close orbits remain close but distinct, corresponding to oscillatory dynamics on a limit-cycle or torus (for tori, at least one other exponent must be zero); and finally for  $\lambda_1 > 0$, small perturbations grow exponentially, and the system evolves chaotically within the folded space of a so-called ``strange attractor'' (for two or more positive definite LEs we speak of ``hyperchaos").
\par
In general, non-linear system equations and the asymptotic nature of the LEs precludes any analytic evaluation. Instead, various methods of numerical approximation of  these quantities, both directly from ODE models and from time-series data\cite{Rosenstein:1993p798, Bryant:1990p429, Gencay:1992p2078} have been developed. In this paper, Lyapunov spectra are calculated using a Python implementation of a method proposed by Benettin {\em et.al.}\cite{Benettin:1980p1676} and Shimada and Nagashima\cite{Shimada:1979p1747}, (outlined in the Supplementary Information online) for inference of Lyapunov exponents when the differential equations are known.
\par
{\color{response}For each of the results presented in this article, we used LSODE to integrate the equations initially for 1000 time steps, in order to overcome the transient dynamics. The Lyapunov exponents were then estimating over a further 10,000 points. The step size varied between $0.01$ for the Lorenz system and $0.5$ for the Hes1 model. The accuracy of our implementation may be gauged from the sum of squared errors shown for the oscillation inference results where the maximal Lyapunov exponent should in theory be $0$. For the Hes1 system, limit cycles attractors were obtained with estimated $|\max(\lambda_i)| < 6\times10^{-3}$, while for the electric ciricuit, the oscillations found had $|\max(\lambda_i)| < 3\times10^{-3}$. Further, for the Lorenz system with typical parameter values $(\sigma, \rho, \beta)=(10, 28, 8/3)$, we estimate the Lyapunov spectrum as $(0.886, -4\times10^{-3}, -15.2)$ as compared to the values $(0.906, 0, -14.57)$ reported by J.C. Sprott\cite{Sprott:2003vc}. In this light, a conservative choice for $\delta_{tol}$, the tolerance level above which we take an estimated maximal Lyapunov exponent to indicate the presence of chaos would be $\approx 0.05$}

\subsection*{Lyapunov spectrum driven parameter estimation}
Unlike in the case for linear systems, where identifying suitable parameters that produce observed or desired dynamics is trivial, inference for highly non-linear systems is far from straightforward. Indeed, exact inferences are prohibitively expensive for even small systems, and so a host of different approximation methods have been proposed\cite{Newton:1994p4219,DelMoral:2009p4056,Toni:2009p9197}. In our case, two further complications arise from using LEs to encode the desired behaviour. Firstly, the form of the mapping between model parameters and LEs is not closed, making methods that rely on an approximation of the estimation routine or its derivatives, such as the extended Kalman filter, difficult to apply. Secondly, LEs are significantly more expensive to compute than more traditional cost functions, ruling out the use of approaches such as particle filtering or sequential Monte-Carlo methods that require extensive sampling of regions of parameter space and calculation of the corresponding LEs at each iteration.
\par
To overcome these challenges we exploit the efficiency and flexibility of the UKF\cite{Wan:2000vj, WanMerwe:2002book, Quach:2007p2717}, seeking here to infer the posterior distribution over parameters that give rise to the desired LEs. Typically the UKF is applied for parameter estimation of a non-linear mapping $g(\cdot)$ from a sequence of noisy measurements, ${ y}_k$, of the true states, $x_k$, at discrete times $k=t_1,..,t_N$. A dynamical state-space model is defined,
\begin{eqnarray}
{ \theta}_k &=& { \theta}_{k-1}+{ v}_{k}\\
{ y}_k &=& {g( x_{k}, \theta}_k)+{ u}_{k}
\end{eqnarray}
where ${ u}_{k-1} \sim N(0, { Q}_k)$ represents the measurement noise, ${ v}_{k-1} \sim N(0, { R}_k)$ is the artificial process noise driving the system, and $g(\cdot)$ is the mapping for which parameters $\theta_k$ are to be inferred. The UKF (described in full below) is then characterised by the iterative application of a two step, {\it predict} and {\it update}, procedure. In the {\it prediction step} the current parameter estimate is perturbed by the driving process noise $ v_{k}$ forming {\it a priori} estimates (which are conditional upon all but the current observation) for the parameter mean and covariance. These we denote as ${ \hat\theta}^{pr}_k$ and ${ P}^{pr}_k$, respectively. The {\it update step} then updates the {\it a priori} statistics using the additional measurement, ${ y}_k$,  to form {\it a posteriori} estimates, ${ \hat\theta}^{po}_k$ and ${ P}^{po}_k$. After all observations have been processed we arrive at the final parameter estimate, ${ \hat\theta}^{po}_{t_N}$ (with covariance ${ P}^{po}_{t_N}$). 
\par
A crucial step in the algorithm is the propagation of the {\it a priori} parameter distribution statistics through the model, $g(\cdot)$. Assuming linearity of this transformation, a closed form optimal filter may be derived (known as the Kalman filter). However, this assumption would make the algorithm inappropriate for use with the highly non-linear systems and the choice of $g(\cdot)$ considered here. It is how the UKF copes with this challenge, namely its use of the ``unscented transform'', that makes it particularly suitable for our method of qualitative feature driven parameter estimation. 
\par
The unscented transform is motivated by the idea that probability distributions are easier to approximate than highly non-linear functions\cite{Julier:1996p2328}. In contrast to the Extended Kalman filter where non-linear state and transition functions are approximated by their linearised forms, the UKF defines a set, $\Theta_k$, of ``sigma-points''  - deterministically sampled particles from the current posterior parameter distribution (given by ${ \hat\theta^{po}_{k-1}}$ and ${ P}^{po}_k$ ), that along with corresponding weights, $\{{\omega^{m}_i,\omega^{c}_i}\}_k$, completely capture its mean and covariance. The mean and covariance of the predicted observation, ${ y}_{k}$, may then be calculated to third order accuracy in the Taylor expansion, using the equations given below. Under the approximate assumption of Gaussian prior and posterior distributions ({\color{response}higher order moments may be captured if desired at the cost of computational efficiency}), the deterministic and minimal sampling scheme at the heart of the filter requires relatively few LE evaluations at each iteration ($2n_p + 1$, where $n_p$ is the number of parameters to be inferred). Further, the function that is the subject of the inference may be highly-nonlinear and can take any parametric form, such as a feed-forward neural network\cite{Wan:2000p5041}, or as in our case, a routine for estimating the Lyapunov exponents of a model with a given parameter set.

With $[X]_i$ denoting the $i^{th}$ column of the matrix $X$, the UKF algorithm for parameter estimation is given by,
\begin{eqnarray*}
\text{\it Initialize:}\\
{\hat \theta}^{po}_0&=&E({ \theta})\\
{ P}^{po}_0&=&E(({ \hat \theta}_0-{ \theta})({ \hat \theta}_0-{ \theta})^T)\\
\text{For each time point $k=t_1,...,t_N$:}\\
\text{\it Prediction step:}\\
{ \hat\theta}^{pr}_k&=&E({ \theta} | { y}_{i\leqslant k-1}) \nonumber \\
			    &=&{ \hat \theta}^{po}_{k-1} \label{tu2}\\		    
\color{response}{ P}^{pr}_k&=&\color{response}{ P}^{po}_{k-1}+{ Q}_{k-1}\\
\text{\it Update step:}\\\\
	{\hat \theta}^{po}_k&=&{ \hat\theta}^{pr}_k+{ K}_k({ y}_k-{ \hat{y}}_k)\\
	{ P}^{po}_k&=&{ P}^{pr}_k-{ K}_k{ P}_{{ \hat{y}}_k}{ K}^T_k\\
\text{where}\\
	{ Y}_{k} &=& g({ x_{k}},{\Theta_k})\\
	{ \hat{y}}_{k} &=& \sum_{i=0}^{2L} \omega^{m}_{i}[{ Y}_k]_i\\
	{ P}_{ \hat{y}_k} &=& \sum_{i=0}^{2L} \omega^{c}_{i}([{ Y}_k]_i-{ \hat{y}}_k)([{ Y}_k]_i-{ \hat{y}}_k)^T+{ R}_{k}\\
	{ P}_{{ \theta}^{pr}_k y_k} &=& \sum_{i=0}^{2L} \omega^{c}_{i}([\Theta_k]_i-{ \hat{\theta}^{pr}}_k)([{ Y}_k]_i-{ \hat{y}}_k)^T\\
	{ K}_k &=& { P}_{{ \theta}^{pr}_{k}{ y}_{k}}{ P}^{-1}_{ \hat{y}_k}
\end{eqnarray*}

Various schemes for sigma-point selection exist including those for minimal set size, higher than third order accuracy and (as defined and used in this study) guaranteed positive-definiteness of the parameter covariance matrices{\color{response}\cite{Julier:1996p2328,Tenne:2003hj, Julier:2002uu, Julier:2002un}, }which is necessary for the square roots obtained by Cholesky decomposition when calculating the sigma-points. The scaled sigma-point scheme thus proceeds as,
$$
\begin{tabular}{l l l l}
$[\Theta_k]_0= { \hat{\theta}}^{pr}_k$ & $$ & $\omega^m_0 = \frac{\lambda}{L+\lambda}$ & $i=0$ \\
$[\Theta_k]_i= { \hat{\theta}}^{pr}_k + \left[{ \sqrt{(L+\lambda){ P}^{pr}_k}}\right]_{i}$  & $i=1,...,L$ & $\omega^c_0 = \frac{\lambda}{L+\lambda}+(1-\alpha^2+\beta)$ & $i=0$\\
$[\Theta_k]_i = { \hat{\theta}}^{pr}_k - \left[{ \sqrt{(L+\lambda){ P}^{pr}_k}}\right]_{i}$ &  $ i=L+1,...,2L$ & $\omega^m_i = \omega^c_i = \frac{1}{2(L+\lambda)}$ & $i=1,...,2L$, \\
\end{tabular}
$$
where $$\lambda = \alpha^2(L+\kappa)-L$$
and parameters $\kappa$, $\alpha$ and $\beta$ may be chosen to control the positive definiteness of covariance matrices, spread of the sigma-points, and error in the kurtosis respectively.
\par
To apply to the UKF for qualitative inference, we amend the dynamical state-space model to,
\begin{eqnarray}
\theta_{k} &=& \theta_{k-1} + v_{k} \\
\lambda_{\text{target}} &=& L(\theta_{k},y_0;f) + u_k,
\end{eqnarray}
where $L(\cdot)$ maps parameters to the encoding of the dynamical behaviour (here a numerical routine to calculate the Lyapunov spectrum), $\lambda_\text{target}$ is a constant target vector of LEs, $y_0$ denotes the initial conditions, and $f$ is the dynamical system under investigation (with unknown parameter vector $\theta$, considered as a hidden state of the system and not subject to temporal dynamics). To see how equations (4) and (5) fit the state-space model format for UKF parameter estimation it is helpful to consider the time series $(\lambda_{\text{target}}, \lambda_{\text{target}}, \lambda_{\text{target}},...)$ as the ``observed'' data from which we learn the parameters of the non-linear mapping $L(\cdot)$. Our use of the UKF is characterised by a repeated comparison of the simulated dynamics for each sigma-point to the {\it same} (as specified) desired dynamical behaviour. In this respect, we use the UKF as a smoother; there is no temporal ordering of the data supplied to the filter since all information about the observed (target) dynamics is given at each iteration. From an optimisation viewpoint, the filter aims to minimise the prediction-error function,

\begin{eqnarray}
E(\theta) = \sum^k_{i=1}[g(\theta,y_0;f) - \lambda_{\text{target}}]^T(Q_k)^{-1}[g(\theta,y_0;f) - \lambda_{\text{target}}],
\end{eqnarray}

\noindent thus moving the parameters towards a set for which the system exhibits the desired dynamical regime.
\par
In the examples presented here, the measurement noise covariance is chosen as $Q_k = aI$, with $a \in {\mathbb R}$ and $I$ the identity matrix, as suggested by E. Wan and R. Van Der Merwe\cite{Wan01chapter7}. We find that varying $a$ over different orders of magnitude does not effect the ability to achieve qualitatively acceptable parameter regimes (results not shown); indeed it can be shown that (if the filter converges) a fixed diagonal measurement noise covariance matrix cancels out of the UKF parameter estimation algorithm\cite{Wan01chapter7}. However, we do find that the choice of $a$ can influence the time taken to reach qualitatively acceptable parameter combinations, with $a \approx 0.01$ performing well for the examples presented here. 
\par
Non-diagonal entries of the process noise covariance, ${P}_k$, are also fixed at zero, with each diagonal entry taken at the same order of magnitude as its corresponding initial model parameter choice. It is worth noting here that unlike in the state-estimation case, the ``artificial'' process noise, $v_{k}$, has no physical interpretation. It may even be set to zero\cite{Quach:2007vv}, though non-zero choices can help the algorithm skip out of non-optimal local minima\cite{Wan01chapter7} and allows the posterior to converge to non-point estimates, or converge at all in the case that the cross-covariance of the parameter prior and predicted data (Lyapunov exponents) remains non-zero. This can be seen by examining the UKF equations for converged ${ P}^{po}_k$. Different methods exist for updating the process noise at each iteration of the filter to control the weight given to past and current observations\cite{Wan01chapter7}, e.g. by annealing the covariance towards zero or by making use of the Robbins-Monro stochastic approximation scheme described in detail elsewhere \cite{Ljung:1983wr}. In the examples presented here, we find that keeping the value fixed gives the most reliable results, possibly reflecting the complex nature of the likelihood surface defined by our choice of $g$.

\subsection*{Hes 1
Quantitative real-time PCR}
Dendritic cells (DC) were differentiated from bone marrow as described previously\cite{Bugeon:2008aa}. 
Rat Jgd1/humanFc fusion protein (R\&D Systems) or human IgG1 (Sigma Aldrich) (control samples) were immobilised onto tissue culture plates (10 $\mu$g/ml in PBS) overnight at 4¡C. DC were spun onto the plate and cells were harvested at the appropriate time. Total RNA was isolated using the Absolutely RNA micro prep kit (Stratagene). cDNA was generated from 125 ng of total RNA using an archive kit (Applied Biosystems). 1 $\mu$l of cDNA was used with PCR Mastermix and TaqMan primer and probes (both Applied Biosystems) and analysed on an Applied Biosystems 7500 PCR system. Cycle thresholds were normalised to 18S and calibrated to a PBS treated control sample for relative quantification. 

\subsection*{Computational Implementation} All routines were implemented in Python using LSODE for integrating differential equations. ABC inference was performed using the ABC-SysBio package\cite{Liepe:2010aa}. Code is available from the authors upon request.

\section*{Acknowledgements}
This research was supported by the UK Biotechnology and Biological Sciences Research Council under the Systems Approaches to Biological Research (SABR) initiative (BB/FO05210/1)  and the {\em Centre for Integrative Systems Biology an Imperial College (CISBIC)} (DS, TT, CB, SM, AR, MJD and MPHS) and a project grant to MPHS; PK was supported by the Wellcome Trust. MPHS is a Royal Society Wolfson Research Fellow.

\section*{Author Contributions}
DS, AR, MJD and  MPHS designed the research; DS, PDWK, CB, TT, AR and SM carried out the research; DS, AR and MPHS wrote the manuscript. All authors have read and approved the final version of the manuscript.

\newpage

\begin{thebibliography}{10}
\expandafter\ifx\csname url\endcsname\relax
  \def\url#1{\texttt{#1}}\fi
\expandafter\ifx\csname urlprefix\endcsname\relax\def\urlprefix{URL }\fi
\providecommand{\bibinfo}[2]{#2}
\providecommand{\eprint}[2][]{\url{#2}}

\bibitem{Tarantola:2005aa}
\bibinfo{author}{Tarantola, A.}
\newblock \emph{\bibinfo{title}{Inverse Problem Theory and Methods for Model
  Selection}} (\bibinfo{publisher}{SIAM}, \bibinfo{year}{2005}).

\bibitem{Moles:2003aa}
\bibinfo{author}{Moles, C.~G.}, \bibinfo{author}{Mendes, P.} \&
  \bibinfo{author}{Banga, J.~R.}
\newblock \bibinfo{title}{Parameter estimation in biochemical pathways: a
  comparison of global optimization methods.}
\newblock \emph{\bibinfo{journal}{Genome Res}} \textbf{\bibinfo{volume}{13}},
  \bibinfo{pages}{2467--2474} (\bibinfo{year}{2003}).

\bibitem{Hartemink:2005aa}
\bibinfo{author}{Hartemink, A.}
\newblock \bibinfo{title}{Reverse engineering gene regulatory networks}.
\newblock \emph{\bibinfo{journal}{Nature Biotechnology}}
  \textbf{\bibinfo{volume}{23}}, \bibinfo{pages}{554--555}
  (\bibinfo{year}{2005}).

\bibitem{Vyshemirsky:2008p14865}
\bibinfo{author}{Vyshemirsky, V.} \& \bibinfo{author}{Girolami, M.~A.}
\newblock \bibinfo{title}{Bayesian ranking of biochemical system models}.
\newblock \emph{\bibinfo{journal}{Bioinformatics}}
  \textbf{\bibinfo{volume}{24}}, \bibinfo{pages}{833--9}
  (\bibinfo{year}{2008}).

\bibitem{Toni:2010p4095}
\bibinfo{author}{Toni, T.} \& \bibinfo{author}{Stumpf, M.}
\newblock \bibinfo{title}{Simulation-based model selection for dynamical
  systems in systems and population biology}.
\newblock \emph{\bibinfo{journal}{Bioinformatics}}
  \textbf{\bibinfo{volume}{26}}, \bibinfo{pages}{104--110}
  (\bibinfo{year}{2010}).

\bibitem{Lu:2006p2418}
\bibinfo{author}{Lu, J.}, \bibinfo{author}{Engl, H.~W.} \&
  \bibinfo{author}{Schuster, P.}
\newblock \bibinfo{title}{Inverse bifurcation analysis: application to simple
  gene systems}.
\newblock \emph{\bibinfo{journal}{Algorithms for molecular biology : AMB}}
  \textbf{\bibinfo{volume}{1}}, \bibinfo{pages}{11} (\bibinfo{year}{2006}).

\bibitem{Khinast:1997vx}
\bibinfo{author}{Khinast, J.} \& \bibinfo{author}{Luss, D.}
\newblock \bibinfo{title}{{Mapping regions with different bifurcation diagrams
  of a reverse‐flow reactor}}.
\newblock \emph{\bibinfo{journal}{AIChE journal}} \textbf{\bibinfo{volume}{43}}
  (\bibinfo{year}{1997}).

\bibitem{Chickarmane:2005p4836}
\bibinfo{author}{Chickarmane, V.}, \bibinfo{author}{Paladugu, S.},
  \bibinfo{author}{Bergmann, F.} \& \bibinfo{author}{Sauro, H.}
\newblock \bibinfo{title}{Bifurcation discovery tool}.
\newblock \emph{\bibinfo{journal}{Bioinformatics}}
  \textbf{\bibinfo{volume}{21}}, \bibinfo{pages}{3688} (\bibinfo{year}{2005}).

\bibitem{Conrad:1984ul}
\bibinfo{author}{Conrad, F.}
\newblock \bibinfo{title}{{Parameter estimation in some diffusion and reaction
  models: an application of bifurcation theory}}.
\newblock \emph{\bibinfo{journal}{Chemical engineering science}}
  (\bibinfo{year}{1984}).

\bibitem{Wu:2004td}
\bibinfo{author}{Wu, X.} \& \bibinfo{author}{Hu, H.}
\newblock \bibinfo{title}{{Parameter estimation only from the symbolic
  sequences generated by chaos system}}.
\newblock \emph{\bibinfo{journal}{Chaos (Woodbury, NY)}}
  (\bibinfo{year}{2004}).

\bibitem{Batt:2007p2419}
\bibinfo{author}{Batt, G.}, \bibinfo{author}{Yordanov, B.},
  \bibinfo{author}{Weiss, R.} \& \bibinfo{author}{Belta, C.}
\newblock \bibinfo{title}{Robustness analysis and tuning of synthetic gene
  networks}.
\newblock \emph{\bibinfo{journal}{Bioinformatics}}
  \textbf{\bibinfo{volume}{23}}, \bibinfo{pages}{2415--22}
  (\bibinfo{year}{2007}).

\bibitem{Rizk:2008p5031}
\bibinfo{author}{Rizk, A.}, \bibinfo{author}{Batt, G.}, \bibinfo{author}{Fages,
  F.} \& \bibinfo{author}{Soliman, S.}
\newblock \bibinfo{title}{On a continuous degree of satisfaction of temporal
  logic formulae with applications to systems biology}.
\newblock In \emph{\bibinfo{booktitle}{Computational Methods in Systems
  Biology}}, \bibinfo{pages}{251--268} (\bibinfo{organization}{Springer},
  \bibinfo{year}{2008}).

\bibitem{Endler:2009p2387}
\bibinfo{author}{Endler, L.} \emph{et~al.}
\newblock \bibinfo{title}{Designing and encoding models for synthetic biology}.
\newblock \emph{\bibinfo{journal}{J R Soc Interface}}
  \textbf{\bibinfo{volume}{6 Suppl 4}}, \bibinfo{pages}{S405--17}
  (\bibinfo{year}{2009}).

\bibitem{GuanrongChen:1997p5902}
\bibinfo{author}{Chen, G.}
\newblock \bibinfo{title}{Control and anticontrol of chaos}.
\newblock \emph{\bibinfo{journal}{Control of Oscillations and Chaos, 1997.
  Proceedings., 1997 1st International Conference}}
  \textbf{\bibinfo{volume}{2}}, \bibinfo{pages}{181 -- 186 vol.2}
  (\bibinfo{year}{1997}).

\bibitem{Tamasevicius:2005p3978}
\bibinfo{author}{Tama{\v s}evi{\v c}ius, A.}, \bibinfo{author}{Mykolaitis, G.}
  \& \bibinfo{author}{Pyragas, V.}
\newblock \bibinfo{title}{A simple chaotic oscillator for educational
  purposes}.
\newblock \emph{\bibinfo{journal}{European journal of {P}hysics}}
  (\bibinfo{year}{2005}).

\bibitem{Momiji:2008p5042}
\bibinfo{author}{Momiji, H.} \& \bibinfo{author}{Monk, N. A.~M.}
\newblock \bibinfo{title}{Dissecting the dynamics of the {H}es1 genetic
  oscillator}.
\newblock \emph{\bibinfo{journal}{J Theor Biol}}
  \textbf{\bibinfo{volume}{254}}, \bibinfo{pages}{784--98}
  (\bibinfo{year}{2008}).

\bibitem{Monk:2003p5886}
\bibinfo{author}{Monk, N. A.~M.}
\newblock \bibinfo{title}{Oscillatory expression of {H}es1, p53, and
  {N}{F}-kappa{B} driven by transcriptional time delays}.
\newblock \emph{\bibinfo{journal}{Curr Biol}} \textbf{\bibinfo{volume}{13}},
  \bibinfo{pages}{1409--13} (\bibinfo{year}{2003}).

\bibitem{Barrio:2006p3389}
\bibinfo{author}{Barrio, M.}, \bibinfo{author}{Burrage, K.},
  \bibinfo{author}{Leier, A.} \& \bibinfo{author}{Tian, T.}
\newblock \bibinfo{title}{Oscillatory regulation of {H}es1: {D}iscrete
  stochastic delay modelling and simulation}.
\newblock \emph{\bibinfo{journal}{PLoS Comput Biol}}
  \textbf{\bibinfo{volume}{2}}, \bibinfo{pages}{e117} (\bibinfo{year}{2006}).

\bibitem{Hirata:2002p4293}
\bibinfo{author}{Hirata, H.} \emph{et~al.}
\newblock \bibinfo{title}{Oscillatory expression of the bhlh factor hes1
  regulated by a negative feedback loop}.
\newblock \emph{\bibinfo{journal}{Science}} \textbf{\bibinfo{volume}{298}},
  \bibinfo{pages}{840--3} (\bibinfo{year}{2002}).

\bibitem{Toni:2009p9197}
\bibinfo{author}{Toni, T.}, \bibinfo{author}{Welch, D.},
  \bibinfo{author}{Strelkowa, N.}, \bibinfo{author}{Ipsen, A.} \&
  \bibinfo{author}{Stumpf, M.~P.}
\newblock \bibinfo{title}{Approximate {B}ayesian computation scheme for
  parameter inference and model selection in dynamical systems}.
\newblock \emph{\bibinfo{journal}{Journal of the Royal Society Interface}}
  \textbf{\bibinfo{volume}{6}}, \bibinfo{pages}{187--202}
  (\bibinfo{year}{2009}).

\bibitem{Kirk:2009aa}
\bibinfo{author}{Kirk, P.} \& \bibinfo{author}{Stumpf, M.}
\newblock \bibinfo{title}{Gaussian process regression bootstrapping: exploring
  the effects of uncertainty in time course data}.
\newblock \emph{\bibinfo{journal}{Bioinformatics}}
  \textbf{\bibinfo{volume}{25}}, \bibinfo{pages}{1300--1306}
  (\bibinfo{year}{2009}).

\bibitem{Castiglione:2008aa}
\bibinfo{author}{Castiglione, P.}, \bibinfo{author}{Falcioni, M.},
  \bibinfo{author}{Lesne, A.} \& \bibinfo{author}{Vulpiani, A.}
\newblock \emph{\bibinfo{title}{Chaos and coarse graining in statistical
  mechanics}} (\bibinfo{publisher}{Cambridge University Press},
  \bibinfo{year}{2008}).

\bibitem{Lorenz:1963p4232}
\bibinfo{author}{Lorenz, E.}
\newblock \bibinfo{title}{Deterministic nonperiodic flow}.
\newblock \emph{\bibinfo{journal}{Atmos. Sci}}  (\bibinfo{year}{1963}).

\bibitem{Qi:2008p586}
\bibinfo{author}{Qi, G.}, \bibinfo{author}{van Wyk, B.} \& \bibinfo{author}{van
  Wyk, M.}
\newblock \bibinfo{title}{Analysis of a new hyperchaotic system with two large
  positive {L}yapunov exponents}.
\newblock \emph{\bibinfo{journal}{Journal of Physics: Conference Series}}
  (\bibinfo{year}{2008}).

\bibitem{Saltelli:2008aa}
\bibinfo{author}{Saltelli, A.}, \bibinfo{author}{Ratto, M.},
  \bibinfo{author}{Andres, T.} \& \bibinfo{author}{Campolong, F.}
\newblock \emph{\bibinfo{title}{Global Sensitivity Analysis: The Primer}}
  (\bibinfo{publisher}{Wiley}, \bibinfo{year}{2008}).

\bibitem{Schroeder:2008aa}
\bibinfo{author}{Schroeder, M.}
\newblock \emph{\bibinfo{title}{Number Theory in Science and Communication}}
  (\bibinfo{publisher}{Springer}, \bibinfo{year}{2008}).

\bibitem{Mitzenmacher:2005aa}
\bibinfo{author}{M.~Mitzenmacher, E.~U.}
\newblock \emph{\bibinfo{title}{Probability and Computing: Randomized
  Algorithms and Probabilistic Analysis}} (\bibinfo{publisher}{Cambridge
  University Press}, \bibinfo{year}{2005}).

\bibitem{Kirk:2008p3388}
\bibinfo{author}{Kirk, P.}, \bibinfo{author}{Toni, T.} \&
  \bibinfo{author}{Stumpf, M.}
\newblock \bibinfo{title}{Parameter inference for biochemical systems that
  undergo a {H}opf bifurcation}.
\newblock \emph{\bibinfo{journal}{Biophysical journal}}
  \textbf{\bibinfo{volume}{95}}, \bibinfo{pages}{540--549}
  (\bibinfo{year}{2008}).

\bibitem{Rosenstein:1993p798}
\bibinfo{author}{Rosenstein, M.}, \bibinfo{author}{Collins, J.} \&
  \bibinfo{author}{Luca, C.~D.}
\newblock \bibinfo{title}{A practical method for calculating largest lyapunov
  exponents from small data sets}.
\newblock \emph{\bibinfo{journal}{Physica D}}  (\bibinfo{year}{1993}).

\bibitem{Bryant:1990p429}
\bibinfo{author}{Bryant, P.}, \bibinfo{author}{Brown, R.} \&
  \bibinfo{author}{Abarbanel, H.}
\newblock \bibinfo{title}{Lyapunov exponents from observed time series}.
\newblock \emph{\bibinfo{journal}{Phys Rev Lett}}
  \textbf{\bibinfo{volume}{65}}, \bibinfo{pages}{1523--1526}
  (\bibinfo{year}{1990}).

\bibitem{Gencay:1992p2078}
\bibinfo{author}{Gencay, R.} \& \bibinfo{author}{Dechert, W.}
\newblock \bibinfo{title}{An algorithm for the n {L}yapunov exponents of an
  n-dimensional unknown dynamical system}.
\newblock \emph{\bibinfo{journal}{Physica D: Nonlinear Phenomena}}
  (\bibinfo{year}{1992}).

\bibitem{Benettin:1980p1676}
\bibinfo{author}{Benettin, G.}, \bibinfo{author}{Galgani, L.},
  \bibinfo{author}{Giorgilli, A.} \& \bibinfo{author}{Strelcyn, J.}
\newblock \bibinfo{title}{Lyapunov characteristic exponents for smooth
  dynamical systems and for hamiltonian systems; a method for computing all of
  them. part 1: Theory}.
\newblock \emph{\bibinfo{journal}{Meccanica}}  (\bibinfo{year}{1980}).

\bibitem{Shimada:1979p1747}
\bibinfo{author}{Shimada, I.} \& \bibinfo{author}{Nagashima, T.}
\newblock \bibinfo{title}{A numerical approach to ergodic problem of
  dissipative dynamical systems}.
\newblock \emph{\bibinfo{journal}{Prog. Theor. Phys}}  (\bibinfo{year}{1979}).

\bibitem{Sprott:2003vc}
\bibinfo{author}{Sprott, J.~C.}
\newblock \emph{\bibinfo{title}{{{C}haos and {T}ime-{S}eries {A}nalysis}}}
  (\bibinfo{publisher}{Oxford Univ Pr}, \bibinfo{year}{2003}).

\bibitem{Newton:1994p4219}
\bibinfo{author}{Newton, M.} \& \bibinfo{author}{Raftery, A.}
\newblock \bibinfo{title}{Approximate {B}ayesian inference with the weighted
  likelihood bootstrap}.
\newblock \emph{\bibinfo{journal}{Journal of the Royal Statistical Society.
  Series B ({M}ethodological)}}  (\bibinfo{year}{1994}).

\bibitem{DelMoral:2009p4056}
\bibinfo{author}{Moral, P.~D.}, \bibinfo{author}{Doucet, A.} \&
  \bibinfo{author}{Jasra, A.}
\newblock \bibinfo{title}{An adaptive sequential {M}onte {C}arlo method for
  approximate {B}ayesian computation}.
\newblock \emph{\bibinfo{journal}{Annals of Applied Statistics}}
  (\bibinfo{year}{2009}).

\bibitem{Wan:2000vj}
\bibinfo{author}{Wan, E.} \& \bibinfo{author}{van~der Merwe, R.}
\newblock \bibinfo{title}{{The unscented Kalman filter for nonlinear
  estimation}}.
\newblock \emph{\bibinfo{journal}{Adaptive Systems for Signal Processing,
  Communications, and Control Symposium 2000.}} \bibinfo{pages}{153--158}
  (\bibinfo{year}{2000}).

\bibitem{WanMerwe:2002book}
\bibinfo{author}{{W}an, E.~A.} \& \bibinfo{author}{van~der {M}erwe, R.}
\newblock \emph{\bibinfo{title}{{K}alman {F}iltering and {N}eural {N}etworks}}
  (\bibinfo{publisher}{{J}ohn {W}iley {\&} {S}ons, {I}nc.},
  \bibinfo{year}{2002}).

\bibitem{Quach:2007p2717}
\bibinfo{author}{Quach, M.}, \bibinfo{author}{Brunel, N.} \&
  \bibinfo{author}{D'Alch{\'e}-Buc, F.}
\newblock \bibinfo{title}{Estimating parameters and hidden variables in
  non-linear state-space models based on odes for biological networks
  inference}.
\newblock \emph{\bibinfo{journal}{Bioinformatics}}  (\bibinfo{year}{2007}).

\bibitem{Julier:1996p2328}
\bibinfo{author}{Julier, S.} \& \bibinfo{author}{Uhlmann, J.}
\newblock \bibinfo{title}{A general method for approximating nonlinear
  transformations of probability distributions}.
\newblock \emph{\bibinfo{journal}{Tech. Rep. RRG, Dept. of Engineering Science,
  University of Oxford}}  (\bibinfo{year}{1996}).

\bibitem{Wan:2000p5041}
\bibinfo{author}{Wan, E.} \& \bibinfo{author}{Merwe, R. V.~D.}
\newblock \bibinfo{title}{The unscented {K}alman filter for nonlinear
  estimation}.
\newblock \emph{\bibinfo{journal}{Adaptive Systems for Signal Processing,
  Communications, and Control Symposium 2000}} \bibinfo{pages}{153--158}
  (\bibinfo{year}{2000}).

\bibitem{Tenne:2003hj}
\bibinfo{author}{Tenne, D.} \& \bibinfo{author}{Singh, T.}
\newblock \bibinfo{title}{{The higher order unscented filter}}.
\newblock \emph{\bibinfo{journal}{Proceedings of the American Control
  Conference}} \textbf{\bibinfo{volume}{3}}, \bibinfo{pages}{2441-- 2446 vol.3}
  (\bibinfo{year}{2003}).

\bibitem{Julier:2002uu}
\bibinfo{author}{Julier, S.}
\newblock \bibinfo{title}{{Reduced sigma point filters for the propagation of
  means and covariances through nonlinear transformations}}.
\newblock \emph{\bibinfo{journal}{Proceedings of the American Control
  Conference}}  (\bibinfo{year}{2002}).

\bibitem{Julier:2002un}
\bibinfo{author}{Julier, S.}
\newblock \bibinfo{title}{{The scaled unscented transformation}}.
\newblock \emph{\bibinfo{journal}{Proceedings of the American Control
  Conference}}  (\bibinfo{year}{2002}).

\bibitem{Wan01chapter7}
\bibinfo{author}{{E}ric {W}an} \& \bibinfo{author}{{R}udolph {V}an~{D}er
  {M}erwe}.
\newblock \bibinfo{title}{{C}hapter 7 {T}he {U}nscented {K}alman {F}ilter}.
\newblock In \emph{\bibinfo{booktitle}{{K}alman {F}iltering and {N}eural
  {N}etworks}}, \bibinfo{pages}{221--280} (\bibinfo{publisher}{Wiley},
  \bibinfo{year}{2001}).

\bibitem{Quach:2007vv}
\bibinfo{author}{Quach, M.}, \bibinfo{author}{Brunel, N.} \&
  \bibinfo{author}{D'Alch{\'e}-Buc, F.}
\newblock \bibinfo{title}{{Estimating parameters and hidden variables in
  non-linear state-space models based on ODEs for biological networks
  inference}}.
\newblock \emph{\bibinfo{journal}{Bioinformatics (Oxford, England)}}
  (\bibinfo{year}{2007}).

\bibitem{Ljung:1983wr}
\bibinfo{author}{Ljung, L.} \& \bibinfo{author}{S{\"o}derstr{\"o}m, T.}
\newblock \bibinfo{title}{{Theory and Practice of Recursive Identification}}.
\newblock \bibinfo{howpublished}{MIT Press, Cambridge, MA}
  (\bibinfo{year}{1983}).

\bibitem{Bugeon:2008aa}
\bibinfo{author}{Bugeon, L.}, \bibinfo{author}{Gardner, L.},
  \bibinfo{author}{Rose, A.}, \bibinfo{author}{Gentle, M.} \&
  \bibinfo{author}{Dallman, M.}
\newblock \bibinfo{title}{Notch signaling induces a distinct cytokine profile
  in dendritic cells that supports {T} cell-mediated regulation and
  {I}{L}-2-dependent {I}{L}-17 production}.
\newblock \emph{\bibinfo{journal}{J. Immunology}}
  \textbf{\bibinfo{volume}{181}}, \bibinfo{pages}{8189--8193}
  (\bibinfo{year}{2008}).

\bibitem{Liepe:2010aa}
\bibinfo{author}{Liepe, J.} \emph{et~al.}
\newblock \bibinfo{title}{{A}{B}{C}-{S}ys{B}io--approximate {B}ayesian
  computation in {P}ython with {G}{P}{U} support}.
\newblock \emph{\bibinfo{journal}{Bioinformatics}}
  \textbf{\bibinfo{volume}{26}}, \bibinfo{pages}{1797--1799}
  (\bibinfo{year}{2010}).

\end{thebibliography}
\appendix
\section*{Appendix}

\subsection*{Lyapunov Exponents}
Lyapunov exponents describe the rate of separation of nearby trajectories and allow the predictability of a system's future states to be quantified (see Figure 1 in the main paper). In our qualitative inference framework we exploit their ability to discriminate between qualitatively different orbit types, allowing us to drive the inference of parameters and initial conditions. Various algorithms exist for the estimation of these quantities both directly from ODE models and from time-series data\cite{Rosenstein:1993p798, Bryant:1990p429, Gencay:1992p2078}.For the results presented here, Lyapunov spectra were calculated using a Python implementation of a method proposed by Benettin {\em et.al.}\cite{Benettin:1980p1676} and Shimada and Nagashima\cite{Shimada:1979p1747}, (outlined below) for inference of Lyapunov exponents when the differential equations are known.
\par

\subsection*{Estimating the Lyapunov spectrum of a differential equation model}

While analytic evaluations of Lyapunov spectra exist for certain special cases or simple systems, generally applicable strategies must resort to numerical estimation techniques. A naive approach to studying a system's sensitivity to initial conditions  would be to directly track the evolution, under $f$, of a set of initially close points and subsequently extract principal rates and directions of contraction/expansion. However, for the following reasons this turns out to be unfeasible:

\begin{itemize}
\item {\color{response}The Lyapunov spectrum describes the average local rates of divergence of nearby trajectories. Under chaotic dynamics and given sufficient time, any deviation from an initial point, no matter how small, will grow too large to represent the local dynamics.}

\item Finite errors in computer calculations and storage mean that every direction in state-space is contaminated by a component in the direction of the dominating $\max(\lambda_i)$. Hence any principal axis evolved through computer simulation will degenerate to align almost entirely along the direction of maximal expansion.
\end{itemize}

A method employing Gram-Schmidt Re-orthonormalization (GSR) addresses these two issues  \cite{Shimada:1979p1747, Benettin:1980p1676} (see Fig. S1). Here an initial $N$-dimensional orthonormal axis, $\{{ e}^0_1,...,{ e}^0_N\}$, with origin at $x_0$, is chosen arbitrarily (e.g. the canonical basis) and integrated simultaneously to the initial condition. Whilst the trajectory from $x_0$ is determined by $f$, each ${ e}^0_i$ is evolved according to $Df({ x}_t)$. The first problem is thus avoided since the linearized equations approximate the real dynamics only at infinitesimally close points to ${ x_t}$. Linearity allows the application of $Df({ x}_t)$ to finite magnitude vectors making up the principal axes. 
\par
The second issue is resolved by a periodic application of GSR to re-orthonormalize the collapsing axes. Let $\{{ e}^t_1,...,{ e}^t_N\}$ be the set of vectors obtained by numerical integration of the principal axes at time $t$; GSR defines a new orthonormal set $\{{ \hat e}^t_1,...,{ \hat e}^t_N\}$ as,

\begin{eqnarray*}
{ \hat e}^t_1 &=& \frac{{ e}^t_1}{||{ e}^t_1||},\\
{ \hat e}^t_2 &=& \frac{{ e}^t_2-\langle{ e}^t_2, { \hat e}^t_1\rangle { \hat e}^t_1}{||{ e}^t_2-\langle{ e}^t_2, { \hat e}^t_1\rangle { \hat e}^t_1||},\\
.\\
.\\
.\\
{ \hat e}^t_N &=& \frac{{ e}^t_N-\langle{ e}^t_N, { \hat e}^t_1\rangle { \hat e}^t_1-...-\langle{ e}^t_N, { \hat e}^t_{N-1}\rangle { \hat e}^t_{N-1}}{||{ e}^t_N-\langle{ e}^t_N, { \hat e}^t_1\rangle { \hat e}^t_1-...-\langle{ e}^t_N, { \hat e}^t_{N-1}\rangle { \hat e}^t_{N-1}||}.
\end{eqnarray*}

Note that the space spanned by the first $k$ vectors is fixed under GSR, and so is free to ``seek'' (driven by numerical error) the $k$-dimensional space with the highest rate of expansion. Thus, since the axes obtained by GSR are orthonormal, the $k$-largest Lyapunov exponents may be obtained from the average rate of growth of the projection of the ${ e}^t_i$ onto ${ \hat e}^t_i$.

\begin{figure}[ht] 
\begin{center}
\includegraphics[width=13cm]{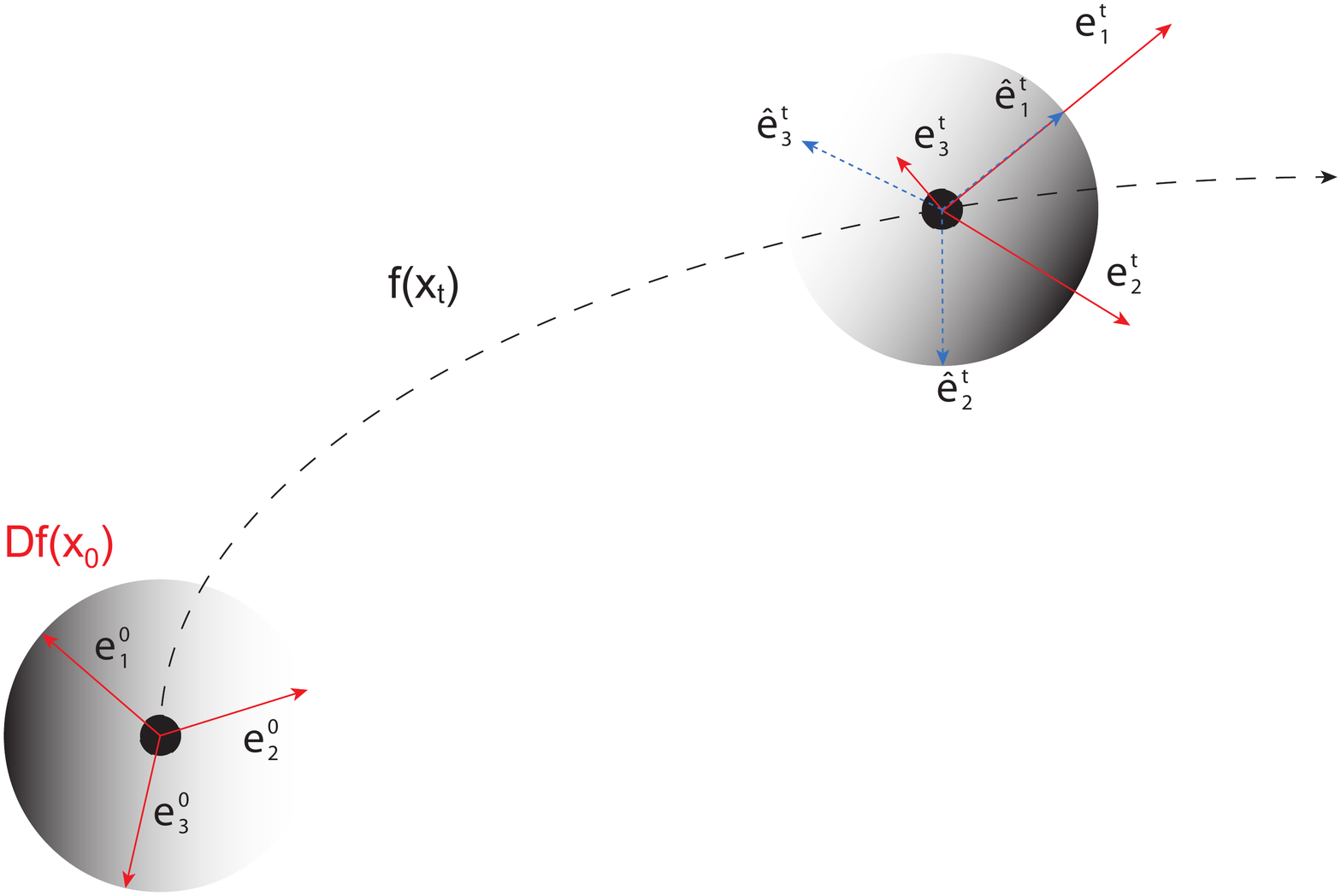}
\caption{\footnotesize{Lyapunov exponents. Lyapunov exponents characterise the long term evolution of the axes of an infinitesimal ball about an initial point in phase space. In the estimation method, an arbitrarily chosen orthonormal axes and the initial condition are evolved simultaneously via the linearised and true system equations respectively. GSR is periodically employed to correct for numerical corruption of the vector directions. Lyapunov exponents are calculated as the average rates of growth of the projections of the evolved axes vectors (blue) onto their re-orthonormalized versions (red).}} 
\label{SupFig1}
\end{center}
\end{figure}

\subsection*{Constraining the inference}

Sometimes we may wish to constrain a search to particular regions of parameter space. In the context of modelling a real world process, this may be based upon the physical impossibility of certain parameter combinations (e.g. negative chemical reaction rates). More generally we may wish to avoid 'badly behaving' regions of parameter space where the model is, for example, unbounded. Instead of constraining the unscented Kalman filter algorithm, we write a new observation function $g^{*}=g\circ p$, where $p$ maps the input parameters onto the region of interest. For example, in order to avoid negative chemical reaction rates, $p$ may output the absolute value of the parameters (and the unchanged model). Note that the parameters inferred by the filter must then be interpreted in light of $p$.

\subsection*{Model equations}

Below we state the governing equations for each of the systems considered. Details of the other models may be found in the references provided in the Results. For each system, a dot above a variable indicates the derivative with respect to time.

\subsection*{ Lorenz map}
\begin{eqnarray*}
\dot{x} &=& \sigma(y-x)\\
\dot{y} &=& x(\rho-z)-y\\
\dot{z} &=& xy-\beta z,
\end{eqnarray*}

\subsection*{ Chaotic electronic circuit}

\begin{eqnarray*}
\dot{x} &=& y\\
\dot{y} &=& ay-x-z\\
\epsilon\dot{z} &=& b+y-c(e^z-1),
\end{eqnarray*}

\subsection*{ Hes1 regulatory model}

\begin{eqnarray*}
\dot{M} &=&-k_{deg}M+1/(1+(P_2/P_0)^{h})\\
\dot{P_1} &=& -k_{deg}P_1+\nu M - k_1P_1\\
\dot{P_2} &=& -k_{deg}P_2+k_1P_1,
\end{eqnarray*}

\subsection*{ Four dimensional hyperchaotic system}

\begin{eqnarray*}
\dot{x_1} &=& a(x_2-x_1)+x_2x_3\\
\dot{x_2} &=& b(x_1+x_2)-x_1x_3\\
\dot{x_3} &=& -cx_3-ex_4+x_1x_2\\
\dot{x_4} &=&  -dx_4+fx_3+x_1x_2
\end{eqnarray*}

\clearpage

\clearpage

\end{document}